\documentclass[fleqn,10pt]{wlscirep}
\usepackage[utf8]{inputenc}
\usepackage[T1]{fontenc}
\usepackage{bm}
\usepackage{lineno}
\usepackage{amsmath}
\newcommand{\angstrom}{\textup{\AA}}
\title{Atomistic deformation mechanism of silicon under laser-driven shock compression}

\author[1,*]{Silvia Pandolfi}
\author[1]{S. Brennan Brown}
\author[2]{P. G. Stubley}
\author[3]{A. Higginbotham}
\author[4]{C. A. Bolme}
\author[1]{H. J Lee}
\author[1]{B. Nagler}
\author[1]{E. Galtier}
\author[4,**]{R. Sandberg}
\author[5]{W. Yang}
\author[6]{W. L. Mao}
\author[2]{J. S. Wark}
\author[1]{A. E. Gleason}

\affil[1]{SLAC National Accelerator Laboratory, 2575 Sand Hill Rd., Menlo Park, CA 94025, USA}
\affil[2]{Department of Physics, Clarendon Laboratory, Univeristy of Oxford, Parks Road, Oxford, OX1 3PU, UK}
\affil[3]{Department of Physics, University of York, Heslington York, YO10 5DD, UK}
\affil[4]{Los Alamos National Laboratory, Los Alamos, New Mexico 87545, USA}
\affil[5]{Center for High Pressure Science and Technology Advanced Research, HPSTAR, Shanghai, China}
\affil[6]{Geological Sciences, Stanford University, 367 Panama St., Stanford CA 94305, USA}
\affil[**]{Now at: Department of Physics and Astronomy, Brigham Young University, Provo, Utah 84602, USA}
\affil[*]{e-mail: silviap@stanford.edu}

\begin{abstract}

Silicon (Si) is one of the most abundant elements on Earth, and it is the most important and widely used semiconductor, constituting the basis of modern electronic devices. Despite extensive study, some properties of Si remain elusive. For example, the behaviour of Si under high pressure, in particular at the ultra-high strain rates characteristic of dynamic compression, has been a matter of debate for decades. A detailed understanding of how Si deforms is crucial for a variety of fields, ranging from planetary science to materials design. Simulations suggest that in Si the shear stress generated during shock compression is released \textit{inelastically}, i.e., via a high-pressure phase transition, challenging the classical picture of relaxation via defect-mediated plasticity. However, experiments at the short timescales characteristic of shock compression are challenging, and direct evidence supporting either deformation mechanism remain elusive. Here, we use sub-picosecond, highly-monochromatic x-ray diffraction to study (100)-oriented single-crystal Si under laser-driven shock compression. We provide the first unambiguous, time-resolved picture of Si deformation at ultra-high strain rates, demonstrating the predicted \textit{inelastic} shear release. Our results resolve the longstanding controversy on silicon deformation under dynamic compression, and provide direct proof of strain rate-dependent deformation mechanisms in a non-metallic system, which is key for the study of planetary-relevant materials.
\end{abstract}

\begin{document}
\flushbottom
\maketitle

\thispagestyle{empty}


The response of materials to ultra-high pressures has been investigated for more than a century \cite{Bridgman1912,Bridgman1914,Bridgman1923,BUNDY1955,Bundy1961,Minomura1962,Lynch1966,Slichter1972,Bundy1980}, fostering our understanding of the fundamental properties of matter as well as the phenomena taking place at the interior of planets within our own solar system and beyond\cite{Bridgman1956,Seager2007,Valencia2009,Swift2011}. Si is one of the most abundant elements in our planet, and it has found wide-ranging application in the semiconductor industry. Because of its technological interest, Si properties have been extensively studied, both at ambient and high-pressure (HP) conditions\cite{Mujica.2003, Wark.1987, Wark.1989}. However, despite decades of research, there is no consensus around the deformation mechanism that drives Si transitions at HP. Precise understanding of these transformation pathways is of interest for fields ranging from planetary science\cite{Duffy.2019} to the recovery of novel functional materials for industrial and energetic applications\cite{Pandolfi.2018,Kurakevych.2016,Kim.2015,Haberl.2016}

Si exhibits a complex phase diagrams, with several HP and metastable polymorphs\cite{Mujica.2003}. Under static loading, upon increasing pressure Si-I transforms into metallic $\beta$-tin Si-II, \textit{Imma} Si-XI, sh Si-V, \textit{Cmca} Si-VI, hcp Si-VII and fcc Si-X\cite{Olijnyk.1984,McMahon.1993,McMahon.1994,Hanfland.1999,Hu.1986,Duclos.1987,Duclos.1990,Anzellini.2019}. Under dynamic loading, Si displays a complex response that has been a matter of debate for decades\cite{Gust.1971,Coleburn.1972,Goto.1982}, with a multi wave profile emerging between 5.4 and 9.2 GPa depending upon crystal orientation and strain rate. Velocimetry experiments have identified the wave following elastic deformation with the onset of a plastic \cite{Smith.2012,Smith.2013} or inelastic\cite{Loveridge-Smith.2001,Turneaure.2007, Higginbotham.2016} regime, followed by a HP phase transition. Molecular dynamic (MD) simulations have suggested that, for dynamic compression of Si, HP phase transitions can occur \textit{inelastically} rather than via defect-mediated plasticity - i.e., the shear stress is released via the HP phase transition, not defect motion \textit{per se}\cite{Mogni.2014}. The activation of a different deformation mechanism at high strain rates could explain the different phase boundaries observed under static-\cite{Mujica.2003} and dynamic-\cite{McBride.2019} compression.

Recent \textit{in situ} x-ray imaging and x-ray diffraction (XRD) of laser-driven Si compression have demonstrated that the emergence of the second wave is due to the onset of the HP phase transition, supporting the inelastic deformation model proposed by MD\cite{McBride.2019, Brown.2019}. However, these experiments could not inform the exact mechanism underlying Si structural transitions, as it requires to establish the specific orientation dependencies between the two phases. 
\textit{In situ} XRD on shock-compressed single-crystal Si using gas gun was performed by Turneaure et al. to study the relative orientation between the ambient Si-I and the HP Si-V phase at 19 GPa\cite{Turneaure.2016}. However, not all the reflections could be indexed in terms of the proposed geometry, preventing an unambiguous interpretation of the experimental data. Thus, the exact deformation mechanism driving Si phase transitions under dynamic compression and the nature of the shear release mechanism (i.e., inelastic vs plastic) remain elusive.

Here, we present a detailed, time-resolved XRD characterization of single-crystal Si(100) under laser-driven shock compression using an ultra-fast X-ray probe from the x-ray free electron laser (XFEL). The use of a highly monochromatic XFEL beam combined with single-crystal starting material ensures high fidelity for the analysis of single-crystal orientation and texture, which is key to investigate Si deformation mechanism. By analyzing the preferred orientation of the HP Si-V phase, we are able to determine the specific transition pathway and to provide the first direct evidence of the inelastic shear release in this compression regime. Comparison with gas gun data resolves previous controversies on the nature of Si deformation, revealing that Si exhibits a complex response to dynamic compression, as it can deform plastically or inelastically depending on the strain rate. 

\begin{figure*}[t]
\includegraphics[width=\textwidth]{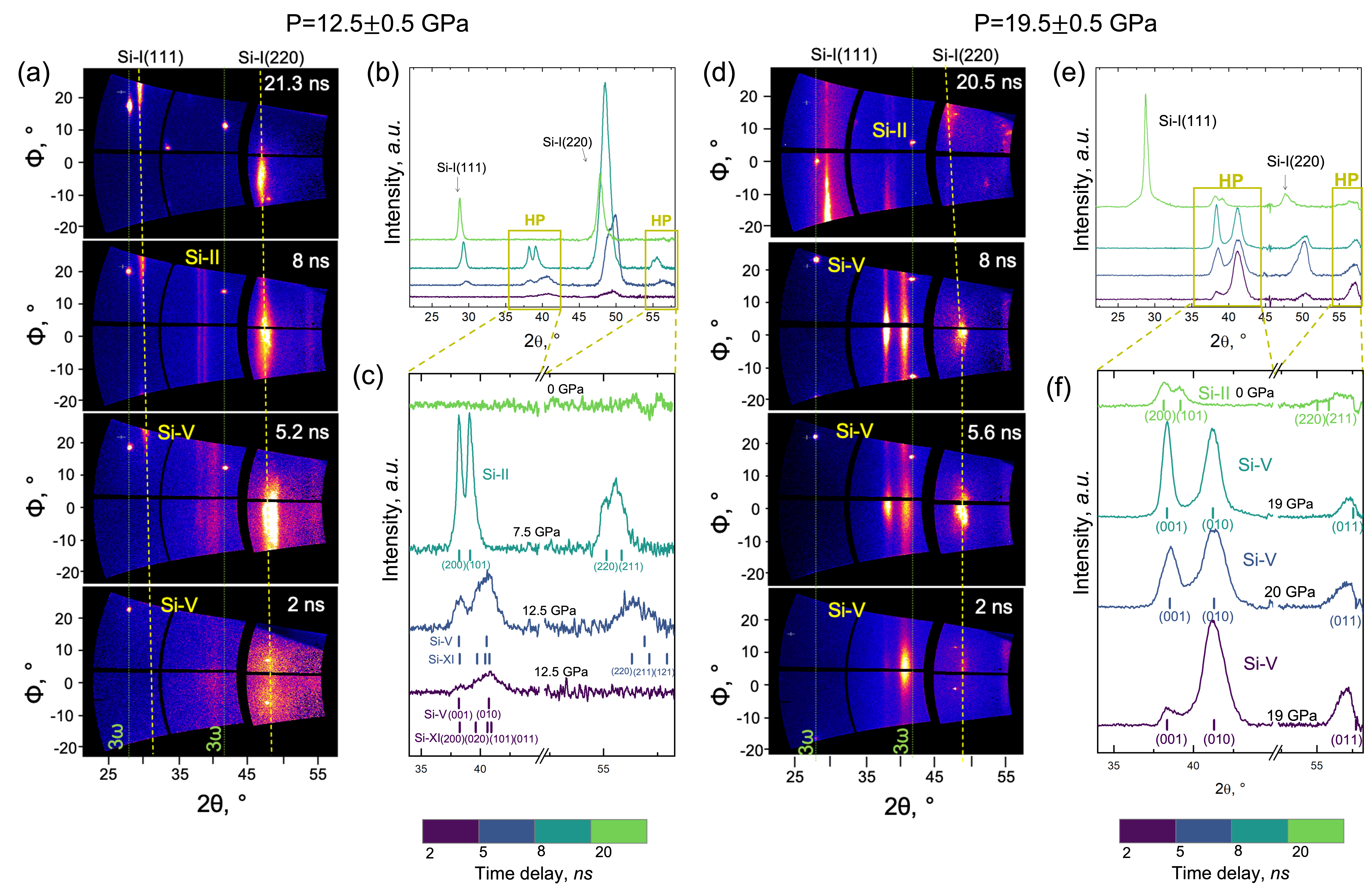}
\caption{XRD data showing the structural evolution of Si(100) during shock compression for two different peak pressures: 12.5$\pm$0.5 GPa (a-c) and 19.5$\pm$0.5 GPa (d-f). (a,d) XRD data projected onto the $2\theta-\phi$ space acquired at different times for compression at 12.5 GPa and 19.5 GPa, respectively. The reflections from Si-I are indexed, and the peak position is indicated by yellow dashed line. The green lines show the position of the third harmonic ($3\omega$) reflections; (b,c,e,f) azimuthally-integrated data are reported with different colors depending on the relative delay between the drive laser and the XFEL pulse; (b),(e) full XRD profile, Si-I peaks are indicated; (c),(f) zoomed view of the high-pressure phases structure; for each pattern the estimated pressure is reported, and the HP phase peaks are indexed.}
\label{fig:fig1}
\end{figure*}

Experiments were conducted at the Matter in Extreme Conditions (MEC) endstation of the Linac Coherent Light Source (LCLS)\cite{Glenzer.2016,Nagler.2015}; single-crystal XRD enabled characterization of the crystal structure and microstructural changes of Si(100) under laser-driven shock compression. XRD data were acquired at time delays of 2, 5, 8 and 20 -ns for two compression regimes, i.e., compression up to 12.5 GPa and 19.5 GPa (Fig.\ref{fig:fig1}). Upon compression, alongside the broad peaks of Si-I, new peaks corresponding to HP phases are visible (Fig.\ref{fig:fig1}(b,e)); these peaks and those of the compressed Si-I were indexed using whole-profile fit. The coexistence of Si-I and HP structures is consistent with MD simulations suggesting a mixed phase region\cite{Mogni.2014} and with recent \textit{in situ} XRD that reported the presence of Si-I above 19.5 GPa and well beyond the Hugoniot elastic limit\cite{McBride.2019,Brown.2019}. The intensity of the XRD signal from the HP phases increases with time, while the peak width decreases, indicating that the HP crystalline domains grow as the shock wave propagates through the material (Fig.\ref{fig:fig1}(c,f)). At 12.5 GPa, shock compression results in a mixed HP phase comprised of both Si-XI and Si-V (Fig.\ref{fig:fig1}(c)) in agreement with previous studies\cite{McBride.2019}. The XRD peaks observed at 2-ns and 5-ns delay fit as both Si-XI and Si-V reflections yielding consistent densities, but a precise deconvolution of the two phases is not possible due to peak superposition. At 19.5 GPa the HP structure is consistent with a pure Si-V phase, even if the asymmetry of the Si-V(010) peaks suggests that small amounts of Si-XI may be present (Fig.\ref{fig:fig1}(f), see also Supplementary Information Sec.5\cite{suppinfo}).

We characterized the structural transitions during both compression and decompression acquiring XRD data up to 20-ns time delay. As inferred from hydrocode simulations (Supplementary Information, Sec.3\cite{suppinfo}), pressure release starts at 5.5 ns for the experiments at 12.5 GPa ($43\mu$m-thick samples), and at 13 ns for those at 19.5 GPa ($100\mu$m-thick samples). During release from 12.5 GPa, we observe the crystallization of Si-II at 8 ns time delay, but the phase is not preserved and it transforms into Si-I upon complete release (Fig.\ref{fig:fig1}(c)). The Si-II formed during decompression has a surprisingly low density, i.e., the measured atomic volume of 15.14 ${\angstrom}^3$/atom would corresponds to a negative pressure value from extrapolation of Si-II Hugoniot curve\cite{Strickson.2016}. This suggests the presence of tensile strain, most likely due to the rarefaction wave generated at the free surface of the sample after 5.5 ns. When releasing from 19.5 GPa, low-density Si-II is observed to be metastable down to ambient pressure, with a measured atomic volume of 15.16 ${\angstrom}^3$/atom (Fig.\ref{fig:fig1}(f)). Despite the ultra-high strain rate, which, according to recent DAC experiments\cite{Lin.2020}, should favour amorphization, we mainly observe Si-I recrystallization upon decompression. Most likely, the crystal-crystal phase transitions are favoured by the high temperature conditions during isentropic pressure release. It is interesting to notice that none of the metastable phases that have been synthesized in static compression experiments (e.g., Si-III\cite{Kurakevych.2016} or a-Si\cite{Lin.2020}) was observed. The extended metastability of metallic Si-II at ultra-high strain rates presents opportunities for future recovery experiments that could extend the capabilities of Si HP synthesis. Indeed, the quench and recovery of HP phases from laser-driven shock compression has been demonstrated for the $\omega$ phase of zirconium\cite{Gorman.2020}; our results suggest that this experimental approach may be applied also for the recovery of metastable metallic Si-II.

\begin{figure}
\centering\includegraphics[width=0.7\columnwidth]{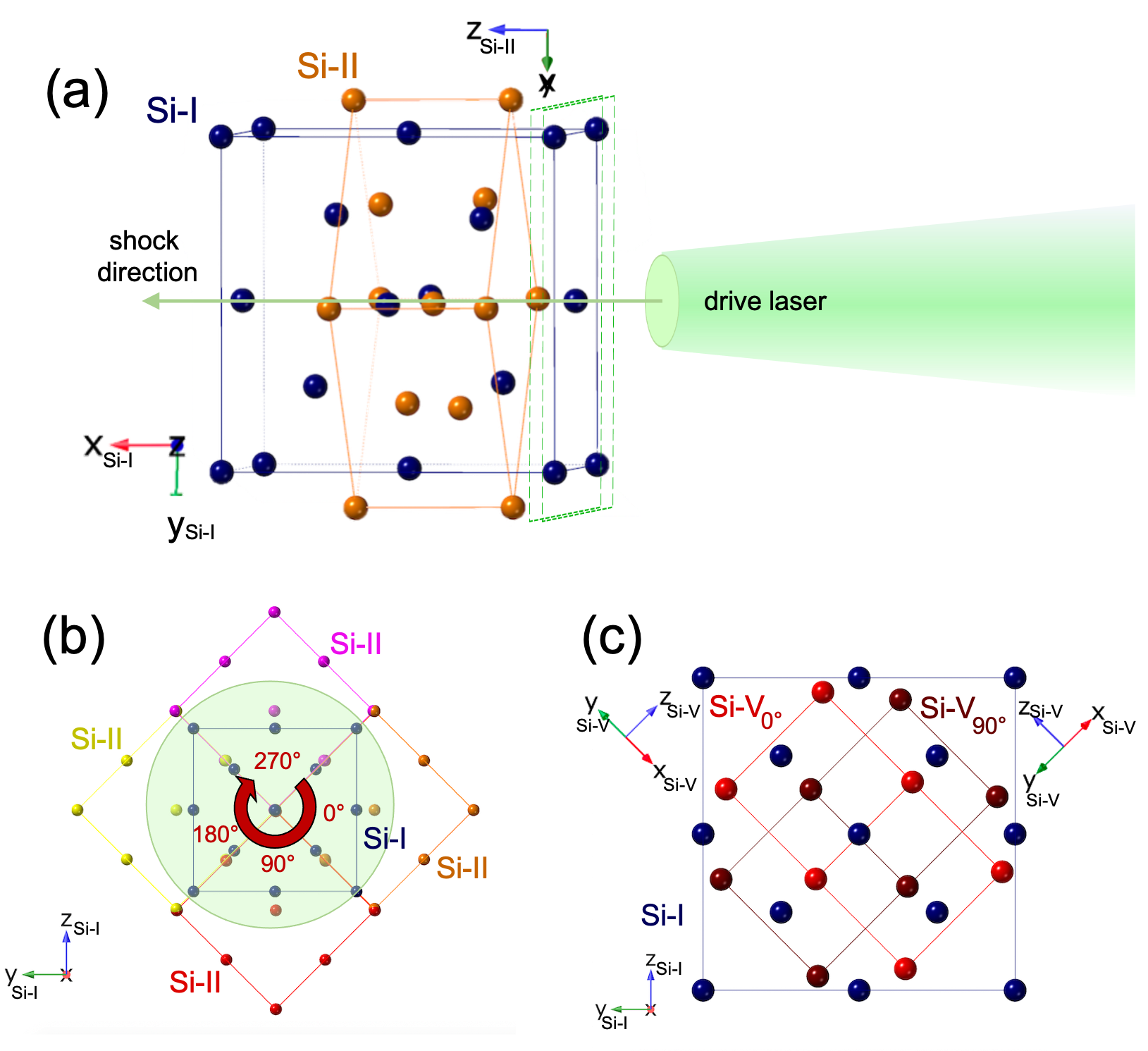}
\caption{Schematic view of the orientation relationship. (a) Preferential orientation for the nucleation of Si-II (orange) from shock-compressed Si-I (blue): $[001]_{II}//[100]_{I}$, i.e,. the compression direction. (b) Perpendicularly to the shock: Si-I (blue) and 4 equivalent Si-II orientations, each of which is indicated with a different color. (c) Orientation relationship between the ambient pressure Si-I (blue) and the HP Si-V phase (red), for which two non-equivalent domains, Si-$V_{0^\circ}$ and Si-$V_{90^\circ}$ are indicated.}
\label{fig:fig2}
\end{figure}

We characterize the relative orientation of the ambient- and high-pressure phases to gain new insight into the atomistic deformation mechanisms that drive Si phase transitions. Indeed, a given deformation mechanism will result in a specific orientation relationship, which can be identified in dynamic compression experiments because shock-compressed materials, being inertially confined, typically exhibit a lower mosaicity compared to static loading. In this study, the use of a highly monochromatic XFEL beam combined with single-crystal starting material ensures an ultra-fast probe for time-resolved study and high fidelity for single-crystal characterization. Upon compression, the XRD signal intensity is not uniform along the Debye-Scherrer rings, confirming the low mosaic spread and suggesting that the HP crystallites grow along preferred directions (Fig.\ref{fig:fig1}(a,d)). In particular, Si-I(220), Si-V(001) and Si-V(010) reflections are centered around the same $\phi$ angle (Fig.\ref{fig:fig1}(d)). The observed relative orientation was reproducible and obtained under different compression regimes, suggesting that it is intrinsic to the transition mechanism (Supplementary Information, Sec.4 \cite{suppinfo}). If dynamic compression resulted in the growth of single-crystal HP phases, Si-V(001) and Si-V(010) reflections would appear at 90$^\circ$ in reciprocal space, as they correspond to perpendicular planes. Thus, Si-V reflections are from two distinct crystalline domains with highly reproducible relative orientation. With the new insights from our XFEL-based experimental approach, we have analyzed several potential orientation relationships, including those proposed by Turneaure et al.\cite{Turneaure.2016}, to explain the highly preferred orientation of Si HP phases under dynamic compression.

\begin{figure}
\centering\includegraphics[width=0.7\columnwidth]{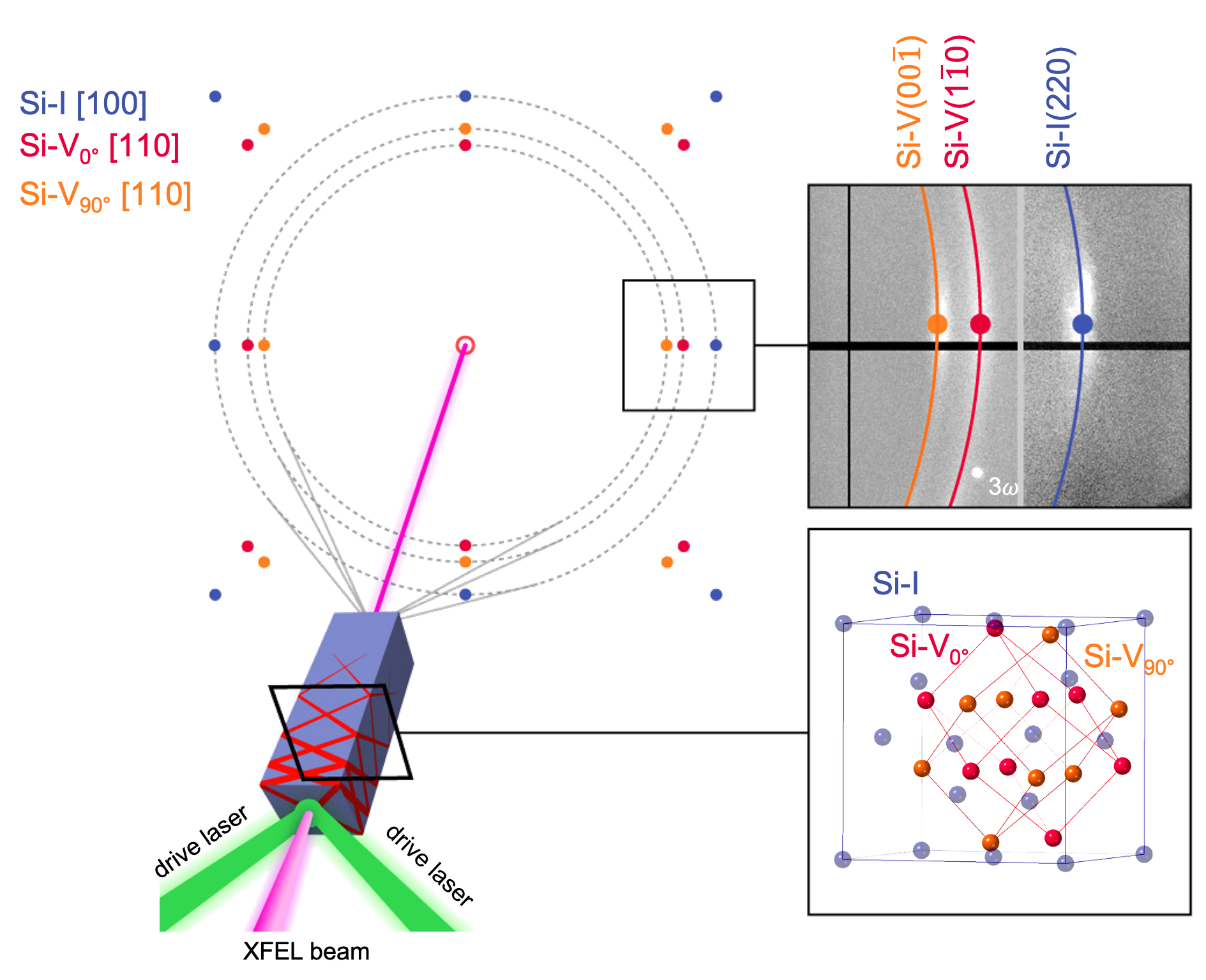}
\caption{Schematic view of our experimental configuration and deformation mechanism. The XRD pattern calculated using the proposed orientation relationship between Si-I and Si-V fits our experimental data (right upper panel).}
\label{fig:fig4}
\end{figure}

We tested the specific orientation relationship dictated by inelastic shear release in Si(100) to verify the deformation mechanism proposed by MD simulations\cite{Mogni.2014}. The initial elastic compression occurs uniaxially along $[100]_I$, i.e., the [100] direction of the Si-I cubic crystal, generating shear stress in the sample. In order for the system to reach the hydrostat inelastically, as part of the sample changes structure to Si-II, the remaining Si-I must experience a compressive stress perpendicularly to the shock in the $(011)_I$ plane.  For the Si-I$\rightarrow$Si-II transition, this is possible if the tetragonal HP phase crystallizes along a specific orientation: $[001]_{II}//[100]_{I}$, i.e., the \textit{c} axis of the HP phase, with shorter inter-atomic distances, is parallel to compression axis (Fig.\ref{fig:fig2}(a)). Perpendicularly to the shock, in the $(011)_I$ plane, Si-II inter-atomic distances are higher, as shown in Fig.\ref{fig:fig2}(b), where the basis vectors of Si-II are rotated through an angle of 45$^\circ$ with respect to the cubic Si-I cell. As a result, if the transition happens with this specific orientation relationship, the relaxation toward the hydrostat (i.e., release of shear stress) is caused by the transition to the highly-oriented HP Si-II phase, and it does not require the generation or motion of crystallographic defects. It is worth noting that this Si-I - Si-II orientation relationship has also been investigated by \textit{ab initio} calculations, and found good agreement with static compression experiments\cite{Biswas.1984}. Because of the four-fold symmetry of the tetragonal phase, there are 4 equivalent crystalline orientations for Si-II in the $(011)_I$ plane (Fig.\ref{fig:fig2}(b)). As Si-II transforms into \textit{Imma} Si-XI and hexagonal Si-V via continuous deformation, the crystal structure is distorted and the four-fold degeneracy is lost ($a_{II}=b_{II} \rightarrow a_{XI}\neq b_{XI}$). The transition to Si-V ultimately results in the formation of crystalline domains with two non-equivalent orientations, as shown in Fig.\ref{fig:fig2}(c): (i) Si-$V_{0^\circ}$, with $[110]_V//[100]_I$ and $[001]_V//[0\overline{1}1]_I$; (ii) Si-$V_{90^\circ}$, with $[110]_V//[100]_I$ and $[001]_V//[011]_I$. The single-crystal XRD pattern calculated using this orientation relationship fits our experimental data well (Fig.\ref{fig:fig4}), while the other geometries we tested could not explain our results. We are thus able to unambiguously interpret our XRD data in terms of a specific deformation mechanism. The sequence of HP phase transitions under uniaxial shock compression results in two non-equivalent Si-V orientations at an angle of 90$^\circ$, which explains the observation of both Si-V(001) and Si-V(100) around the same $\phi$.

\begin{figure}
\centering\includegraphics[width=0.6\columnwidth]{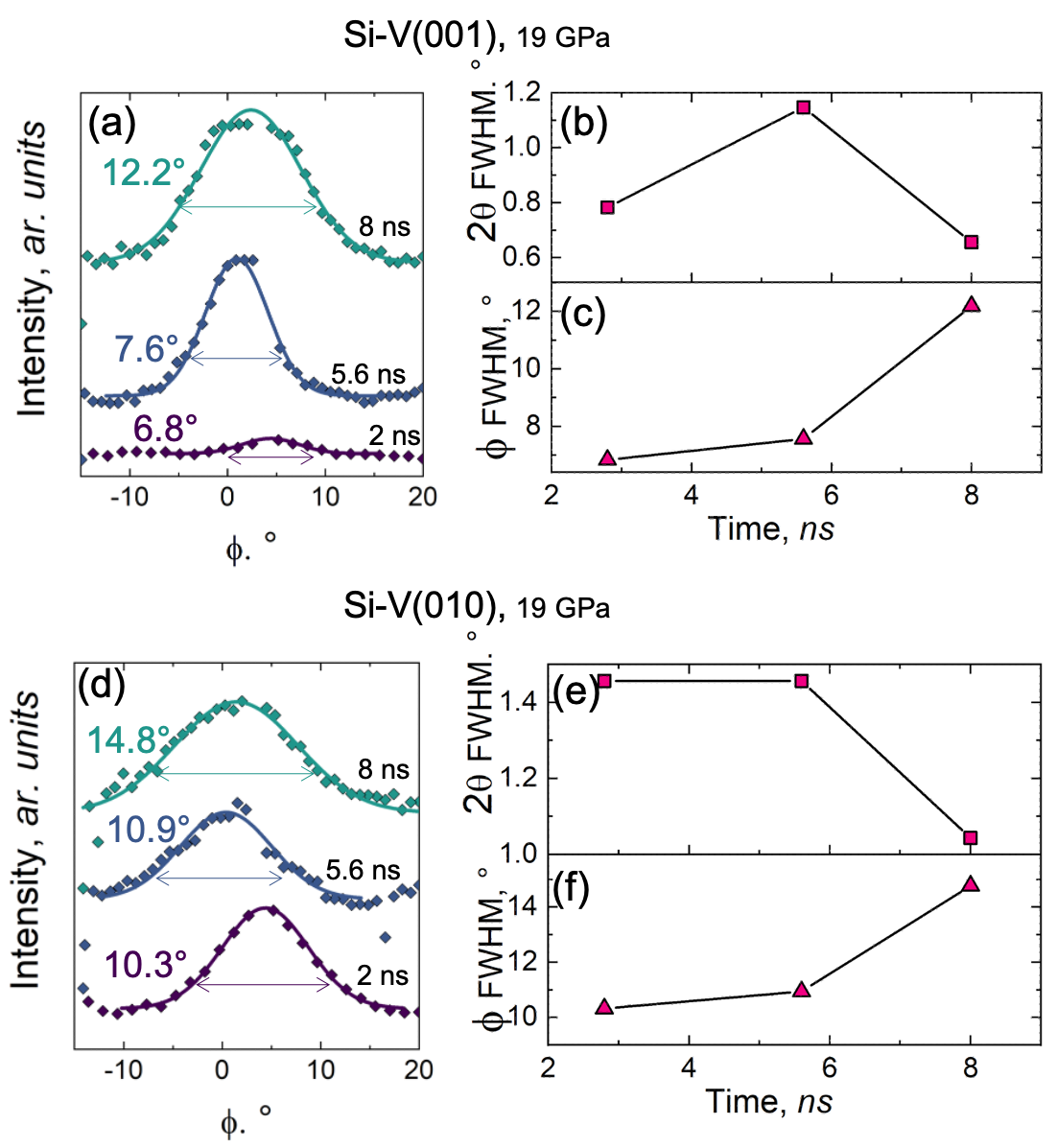}
\caption{Microstructure analysis. (a,d) $\phi$ lineouts for Si-V(001) and Si-V(01) peaks at different times. Experimental data are reported and overlapped with peak fitting results (continuous line and FWHM values form fit).(b,c,e,f) time evolution of peak width along $2\theta$ and $\phi$ for both analyzed peaks.}
\label{fig:fig3}
\end{figure}

We analyzed the XRD peak shapes to gain insight on the microstructure of shock-compressed Si, e.g., changes in mosaicity, grain size, crystallographic defects. The Si-V(001) and Si-V(010) reflections observed around the same $\phi$ value necessarily come from two distinct crystal domains, which prevents analysis of the microstructure via whole-profile refinement (e.g., Le Bail); therefore, we used single-peak fitting projecting the XRD spots along both the $2\theta$ and $\phi$ axis (Fig.\ref{fig:fig3}). The Full Width at Half Maximum (FWHM) along $\phi$ gives an estimate of the disorder of and within the crystalline domains, i.e., a combination of the mosaicity (the angular spread of the orientations of the crystallites\cite{Bellamy.2000}) and rotation of the lattice planes\cite{Heighway.2021} (Fig.\ref{fig:fig3}(a,d)). The $\phi$ lineouts of Si-V peaks at 19.5 GPa show that the peak width along $\phi$ increases with time (Fig.\ref{fig:fig3}(c,f)) in conjunction with the growth of the HP crystallites, i.e., with the decrease of the $2\theta$ FWHM (Fig.\ref{fig:fig3}(b,e)). At 12.5 GPa we observe a similar trend, and the mosaicity increase results in a powder-like texture when Si-II is formed during decompression (Fig.\ref{fig:fig1}(a)). In this regime, the XRD peaks are generally broader and less intense, preventing a quantitative analysis of mosaicity. XRD analysis suggests thus nucleation of highly-oriented HP crystalline domains, with a gradual loss of preferred orientation as the crystalline domains grow. These trends are consistent with the model of HP nano-crystalline domains originating from shear bands along specific crystalline direction, e.g., the Si-I[111] suggested by MD\cite{Mogni.2014}. It is worth noting that previous \textit{ex situ} characterization of shock-compressed Si(100) also showed that amorphous bands originate along \{111\} slip planes, and, as they grow, they tend to deviate from this orientation and align with the direction of maximal shear (i.e.. 35.3$^\circ$ with shock direction in Si(100))\cite{Zhao.2016,Zhao.2015}.

\begin{figure}
    \centering
    \includegraphics[width=0.6\columnwidth]{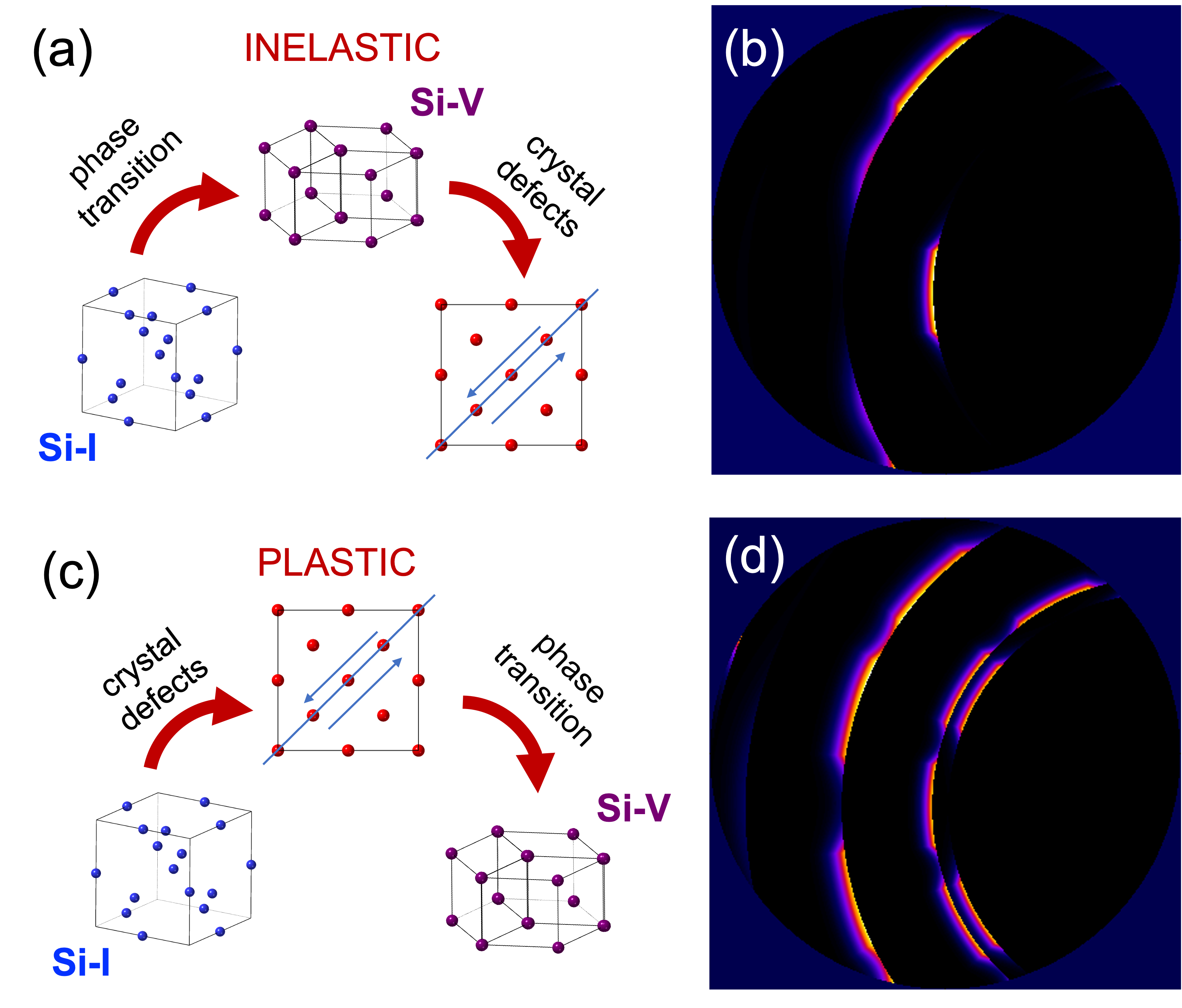}
    \caption{Analysis of the inelastic and plastic deformation regimes, and their XRD signature. (a,c) schematic view of the two deformation mechanisms. (b,d) corresponding XRD signature from forward calculations using the orientation relationship proposed in this work and a pink beam (Supplementary Information, Sec.6\cite{suppinfo})}
    \label{fig:fig5}
\end{figure}

Comparison with previous gas-gun experiments provide additional insights into Si deformation under dynamic compression. The differences between our results and previous experiments by Turneaure et al.\cite{Turneaure.2016} could be due to the specifics of the gas gun experiment, e.g., the use of a pink beam, or by the lower strain rate of gas-gun compression, that may activate different deformation mechanisms\cite{Higginbotham.2016}. In order to perform a direct comparison with the data by Turneaure et al., we performed forward XRD calculations using the orientation relationship proposed in this work and the experimental parameters of previous gas-gun studies (for details, see Supplementary Information, Sect.6\cite{suppinfo}). XRD patterns were simulated for both inelastic and plastic deformation. As shown by the schematics in Fig.\ref{fig:fig5}(a), inelastic deformation is characterized by shear release via a phase transition, followed by the generation of defects and texture evolution; the growth of the HP phase happens only along a specific orientation, resulting in highly localized XRD signal (Fig.\ref{fig:fig5}(b)). Instead, when the sample deforms plastically, the system relaxes toward the hydrostat via generation and motion of defects before the phase transitions take place (Fig.\ref{fig:fig5}(c)). As the compression is not uniaxial, the HP phase can crystallize isotropically, which results in the additional visible reflections shown in Fig.\ref{fig:fig5}(d), where the XRD signal was calculated including all the equivalent variants of the given orientation relationship. Thus, our calculations demonstrate that inelastic and plastic deformation result in different crystalline geometries, with clearly distinguishable XRD signatures, which explains the differences between our results and previous XRD studies\cite{Turneaure.2016}. At lower strain rates, like in gas gun experiments, shock-compression results in defect-mediated plasticity, while the inelastic deformation characteristic of laser-ablation experiments, induces a strongly preferred orientation for the growth of the HP phases. With the new insight provided by our time-resolved single-crystal analysis, we can explain the discrepancies in previous studies on shock-compressed Si, resolving a long standing debate.

In this study, we propose the first unified model for Si deformation under shock-compression, which has been the matter of vigorous debate for decades. Time-resolved XRD is used to characterize, \textit{in situ}, the phase transitions of single-crystal Si(100) under laser-driven shock compression.  With this experimental approach, we provide the first unambiguous, atomistic picture of the deformation mechanism driving phase transitions under laser-driven compression. We observe crystallization of the Si-V HP phase above 12 GPa, and provide evidence of extended metastability of metallic Si-II upon release. The use of a monochromatic XFEL probe with sub-picosend temporal resolution allows us to determine the orientation relationship between the Si-V and Si-I phases, connecting it with the specific deformation mechanism and confirming the inelastic shear release predicted by MD simulations\cite{Mogni.2014}. Analysis of the microstructural evolution demonstrates that at ultra-high strain rates, the HP Si phases form along specific directions with a loss of preferred orientation and generation of defects during crystal growth. Crucially, comparison with previous gas gun experiments demonstrates that at different strain rates the deformation mechanism and the geometry of the phase transition can be substantially different, resolving the ongoing debate around shock-compressed Si.

\section*{Methods}
Single-crystal, double-side parallel-polished platelets of Si(100) (either 43$\mu$m or 100$\mu$m thick) were mounted such that the starting orientation with respect to the probe direction and the shock propagation direction could be tracked. The LCLS delivered 60-fs duration quasi-monochromatic X-ray pulses of energy 7.952 keV ($\Delta{E}$ = 15-40 eV\cite{Emma.2010}, $\Delta{E}/E=0.2–0.5\%$). The XRD signal was recorded on two Cornell-SLAC Pixel Array Detectors covering range of scattering angle $20^\circ < 2{\theta} < 70^\circ$ and azimuth $-25^\circ<\phi<25^\circ$\cite{Nagler.2015}. In this experimental geometry, no XRD signal from the single-crystal starting material is observed for the monochromatic 7.952 keV X-ray probe; spots from Si-I are visible only from crystallites in appropriate orientation and upon compression, when mosaicity and strain increase. However, additional spots due to the parasitic third-harmonic beam are always visible and allow us to evaluate changes in the orientation of the starting material. Shock compression was achieved by direct irradiation of the sample by a 527-nm laser with a quasi-flattop pulse profile of 10-ns duration \cite{Brown.2017} with ~1$\times10^{12}$ $W/{cm}^{2}$ intensity. We collected data at time delays up to 20 ns, which enabled characterization of the transformations taking place during both compression and release, i.e., after the shock wave has traversed the sample. Pressure was estimated from the experimental density obtained from XRD and the multi-phase equation of state of Si\cite{Strickson.2016}. Velocimetry data obtained using the velocimetry interferometry system for any reflector (VISAR) diagnostic and hydrocode simulations were used to confirm pressure and strain-rate estimation (Supplementary Information Sec.2-3\cite{suppinfo}). By varying the time delay between the drive laser and the XFEL beam, time-resolved XRD data were collected at different pressures. For more details, see Supplementary Information Sec.1\cite{suppinfo}.

\bibliography{bibliography.bib}
\section*{Acknowledgements}
This work was performed at the Matter at Extreme Conditions(MEC) instrument of LCLS, supported by the US DOE Office of Science, Fusion EnergyScience under contract No. SF00515, and was supported by LCLS, a National UserFacility operated by Stanford University on behalf of DOE-BES. AEG and WLM are supported in part by the Geophysics Program at NSF (EAR0738873) with additional support from LANL Reines LDRD for AEG. AEG and SP also acknowledge support from 2019 DOE FES ECA. JSW is grateful for support from the UK EPSRC under grant number EP/S025065/1. 





\end{document}